\documentclass[twocolumn,showpacs,preprintnumbers,amsmath,amssymb,superscriptaddress]{revtex4}

\usepackage{graphicx}
\usepackage{dcolumn}
\usepackage{bm}

\begin{document}

\preprint{CAB-lcasat/04021 }


\title{Reaction-diffusion model for pattern formation in $E. coli$ swarming
  colonies with slime.}

\author{M.-P. Zorzano }
\email{zorzanomm@inta.es}
\homepage{http://www.cab.inta.es}

\author{D. Hochberg}%

\author{M.-T. Cuevas}%

\author{J.-M. G\'{o}mez-G\'{o}mez}

\affiliation{Centro de Astrobiolog\'{\i}a (CSIC-INTA),
        Carretera de Ajalvir km 4,    28850 Torrej\'{o}n de Ardoz, Madrid, Spain}     

\date{\today}

\begin{abstract}

A new experimental colonial pattern and pattern transition observed in
$E. coli$ MG1655 swarming cells grown on semi-solid agar are
described. We present a reaction-diffusion model that, taking into
account the slime generated by these cells and its influence on the
bacterial differentiation and motion, reproduces the pattern and
successfully predicts the observed changes when the colonial collective motility is
limited. In spite of having small non-hyperflagellated swarming
cells, under these experimental conditions {\sl E. coli}
MG1655 can very rapidly colonize a surface, with a low branching rate, thanks to a strong
fluid production and a local incremented density of motile,
 lubricating cells.

\end{abstract}


\maketitle


\section{Introduction}

The macroscopic pattern exhibited by a bacterial colony and its
dependence on certain environmental parameters can give us
some clues about the coordinated colonization strategy followed by the
  community of cells. The biological
interest in this interdisciplinary area is in pointing out under controlled laboratory conditions the
cooperative mechanisms (intercellular interactions, motility and
communications) that these bacteria, which are traditionally considered as
solitary life forms, may have developed to adapt
to changing environments. It is an extremely fertile and interesting area
for close collaboration between physicists and biologists
because it helps both to understand the transition
from individual (uni-cellular, a bacteria) to collective (multi-cellular, a colony)
behavior \cite{Shapiro3}.

 Bacterial colony pattern formation on semi-solid
  agar surfaces has been studied extensively by microbiologists and
    physicists \cite{Shapiro3,Viseck,Harshey,serratia, proteus,
  Burkart, phys, Kearns}. Through these works undertaken with
    colonies arising from different species and strains of bacteria, the following common features have been described: Bacteria can swim in liquid medium without difficulty but in environments with
     adverse conditions
     for the swimming motility, they need to develop
    mechanisms to become more motile. On semi-solid agar
    medium, where the viscosity is high and the motility of short swimmer cells
very low, some may differentiate into very elongated, 
 multinucleated and profusely
 flagellated swarm cells that can move easier \cite{Harshey}.  The initiation of swarm
    differentiation seems to be strictly correlated with physico-chemical factors such as
    surface contact and 
    quorum sensing response (cell density sensing mechanism) \cite{serratia}.  Swarming cells
    have the ability to extract water from the agar and
produce a lubrication fluid (slime). The flagellum driven motility of swarmer cells together with the
    extracellular slime helps to overcome the surface friction \cite{Harshey}.  After some time
migrating swarmer cells have been observed to cease movement, septate and
produce groups of swimmer cells (de-differentiation process) \cite{proteus}. The chemotaxis system is essential for
    swarming motility in {\sl E. coli} \cite{Burkart}. With the chemotaxis mechanism bacteria can orient
    their motion in response to a gradient (positive or negative) of
a certain chemical field (a nutrient or a field produced by the
 bacterial cells such as 
 chemical signals or a pH change) \cite{phys}. Based on these experimental observations great effort has been devoted to
model the cooperative behavior of a cell community \cite{Bjacob0, Matsu2, Kawasaki, Golding,
 Bjacob}. 

Here we describe  the new pattern and pattern transitions
   observed in a swarming colony of {\sl E. coli} MG1655
   \cite{blatner}, a laboratory domesticated
   wild type strain that produces a lubricating fluid or slime: when the
   colonial motility is limited 
   this colony expands irregularly with few relatively thick, dense, branches
   and very rare secondary branches, whereas for conditions of improved
   colonial motility the pattern is compact and round with structures of
   higher cell densities. To study the coordinated, self-organized, colonization
   scheme of this colony and based on 
  experimental microscopic observations, we will define a mathematical model of the cooperative behavior of this colony which is able to predict
  the pattern transitions when certain control parameters are changed. 

This work is organized as follows: in Section \ref{exper} we will
   introduce the experimental results and their
implications for the non-expert reader. Then in Section \ref{model}, our hypothesis for the
collective dynamics will be explained, followed by a mathematical
formulation. Results of the computer simulation in comparisons with
   experiments will be posed and
discussed in Section \ref{val}. The summary and conclusions are given
   in Section \ref{conclu}.



\section{Experimental observations}\label{exper}

\begin{figure}
\begin{tabular}{cc}
 (a) & (b)\\
\includegraphics[width=.25 \textwidth]{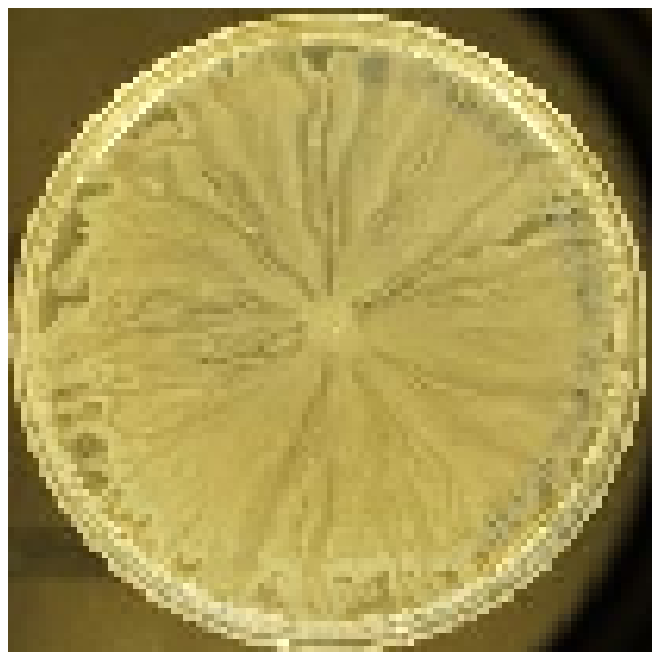}&
\includegraphics[width=.25 \textwidth]{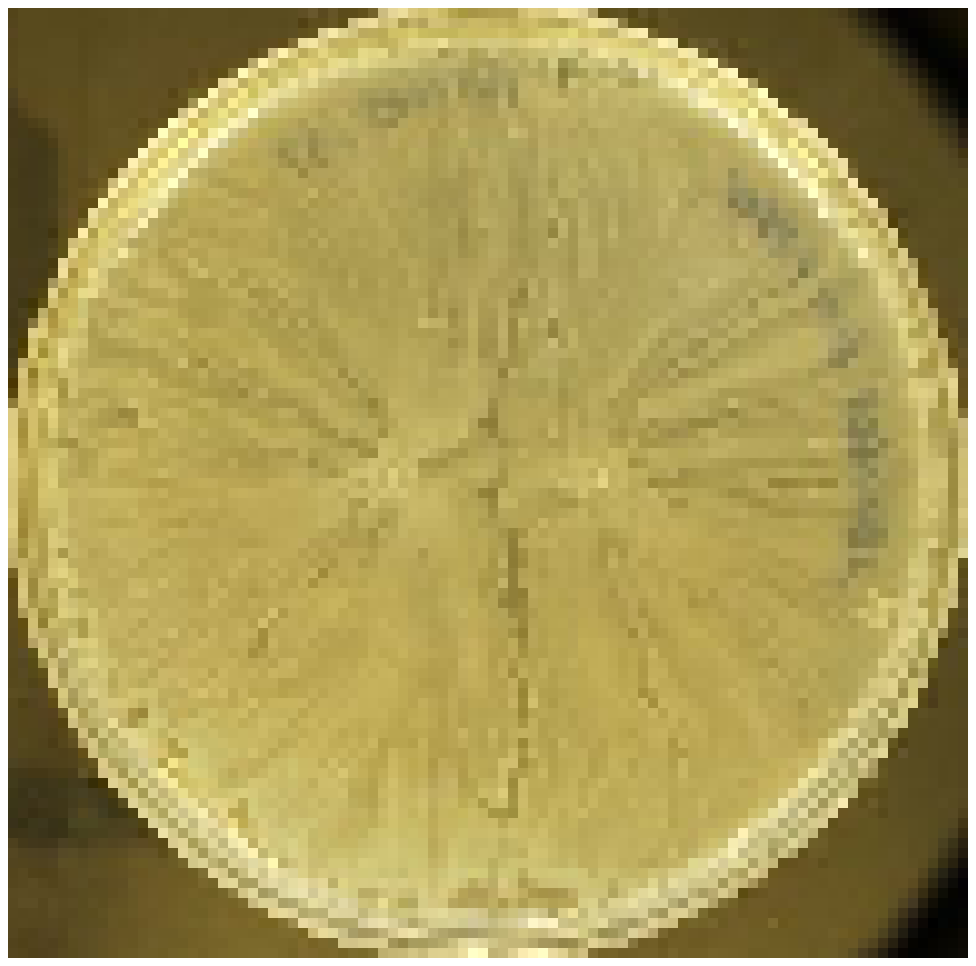}\\
  (c) & (d) \\
\includegraphics[width=.25 \textwidth]{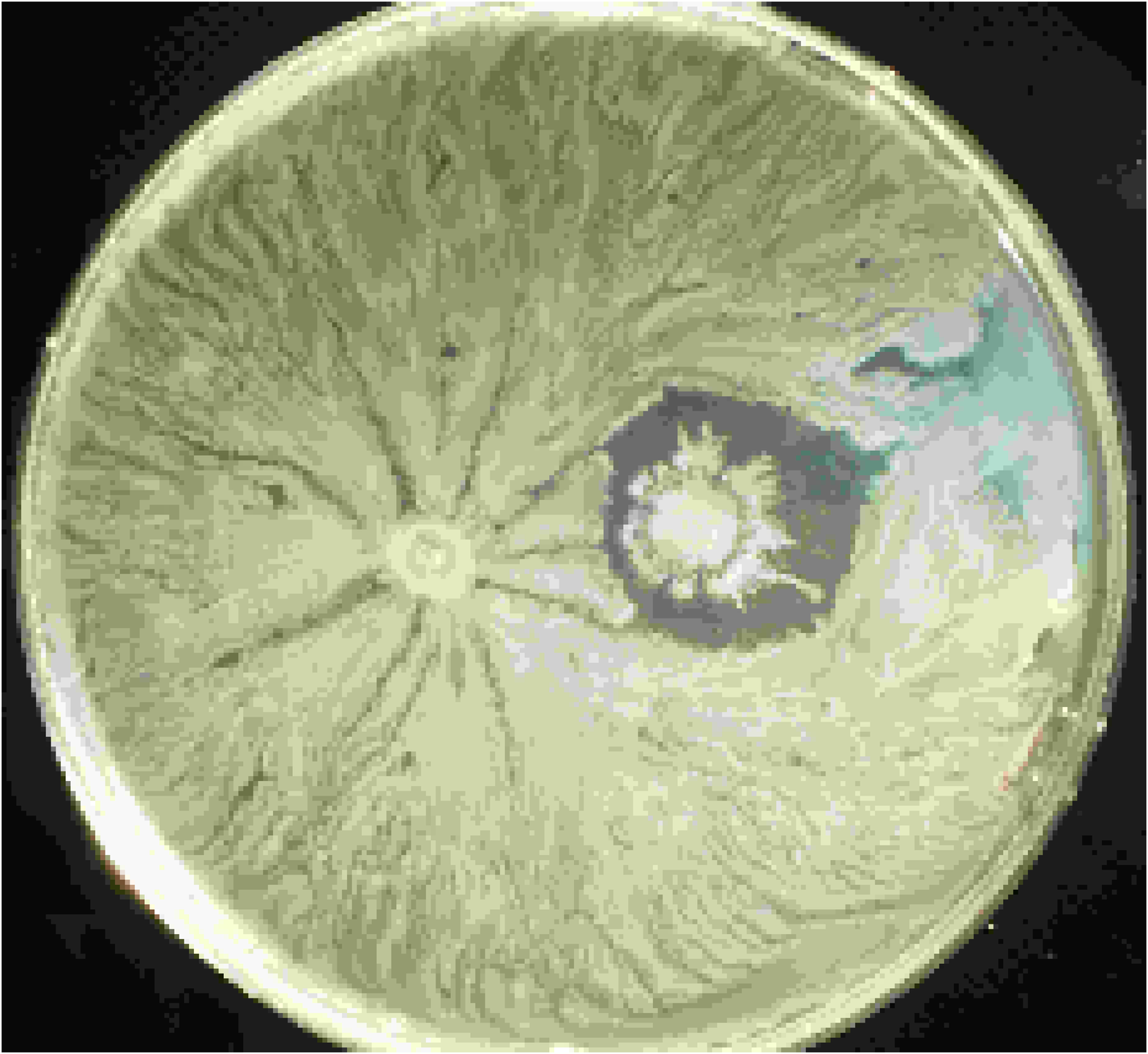}&
\includegraphics[width=.25 \textwidth]{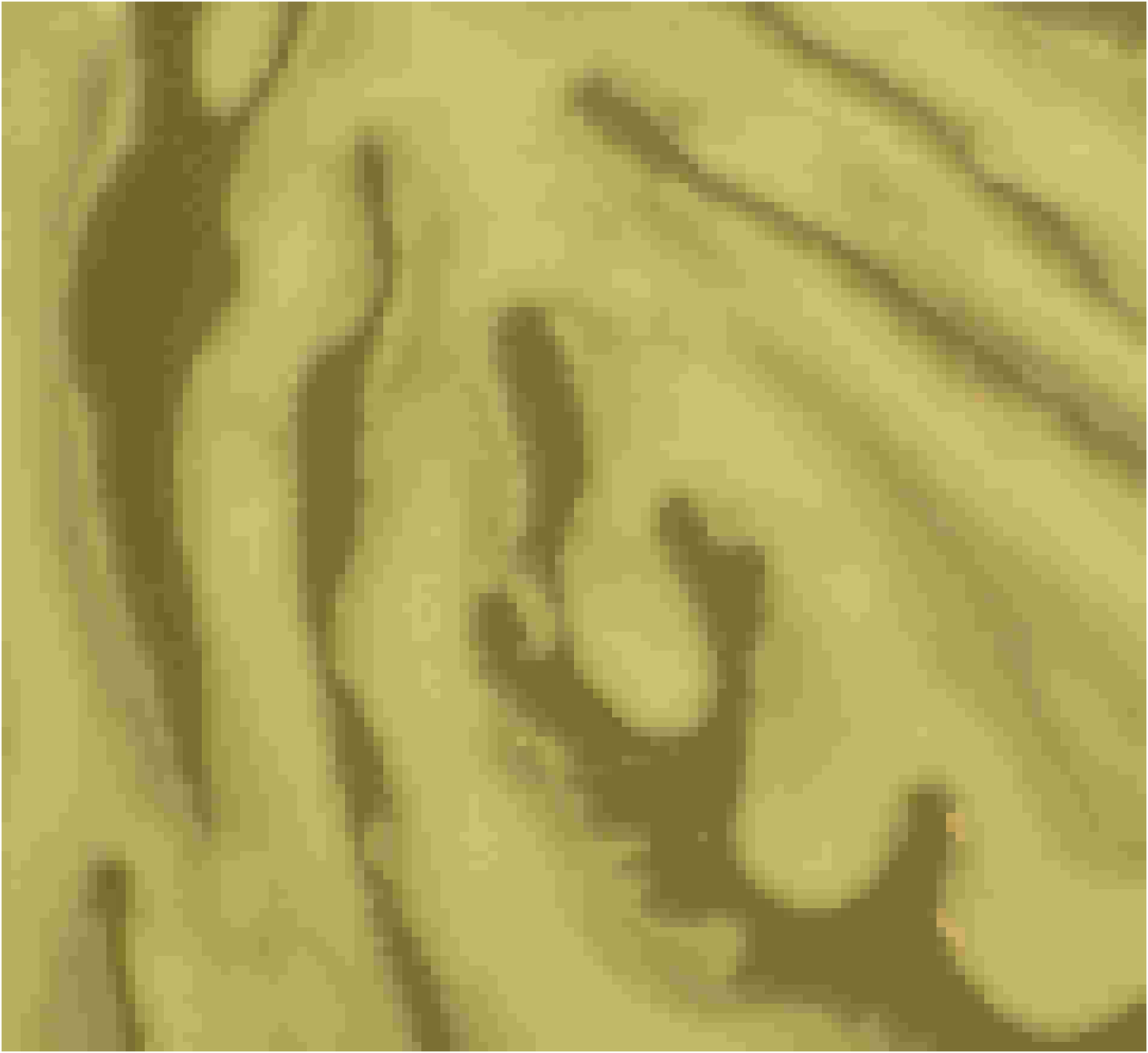}
\end{tabular}
 \caption{\label{Ecoliflor} (a) (Color online) Surface colonial pattern observed in {\sl E. coli}
   MG1655 strain in semi-solid agar. (b) (Color online) Two colonies grown on the
   same plate avoid each other. (c)  (Color online) Comparison of the swarming
 flagellar-dependent surface colonization of the wild type MG1655 inoculated on the left, with
 the spreading growth of the flagella-defective mutant MG1655
  {\sl flhD}::Kan inoculated on
 the right. The fast advancing swarming colony avoids the slowly
 growing one leaving an intercolonial demarcation region. (d) (Color online) Close
 view (box of 2 cm side) of a ramified pattern when the experimental
 conditions are changed (grown at lower temperature $30^{\circ} $C) showing relatively thick branches with increased cell densities.}
\end{figure}

\begin{figure}
\begin{tabular}{cc}
(a) & (b) \\
\includegraphics[width=.25 \textwidth]{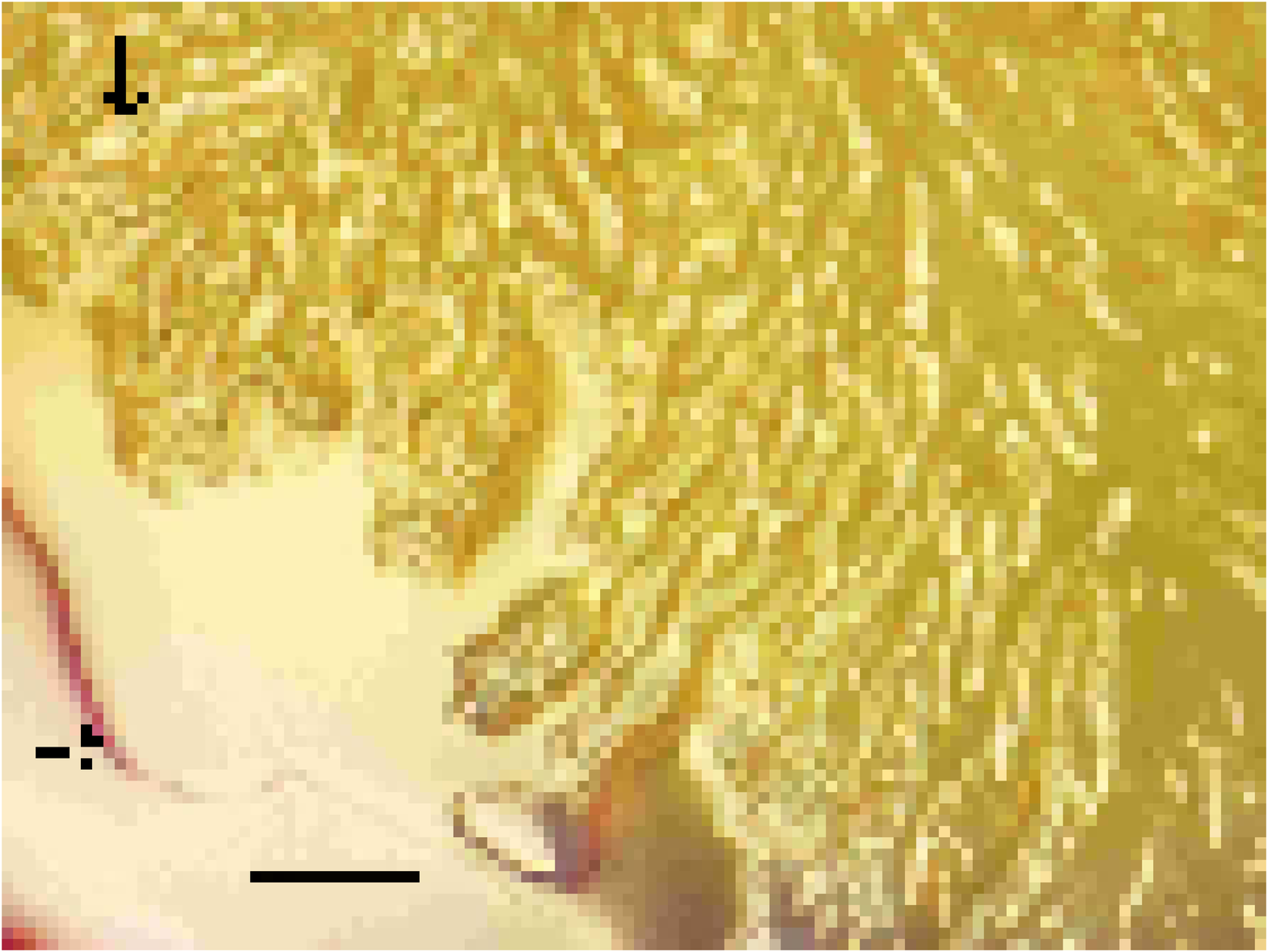}&
\includegraphics[width=.25 \textwidth]{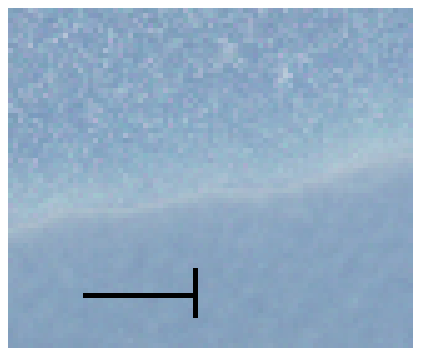}
\end{tabular}
\caption{\label{Otros_Ecoliflor} (a)  (Color online) Close view of colony edge with 
 surrounding slime (lower arrow) and channel-like
   structure (upper arrow). (Bar=$5\ $mm). (b) (Color online) Swarming cells at the edge of
   the colony, bounded with a thin layer of quiescent and packed swarming
   cells. (Bar=$50\ \mu $m).}
\end{figure}

\begin{figure}
\begin{tabular}{cc}
 (a) & (b) \\
\includegraphics[width=.25 \textwidth]{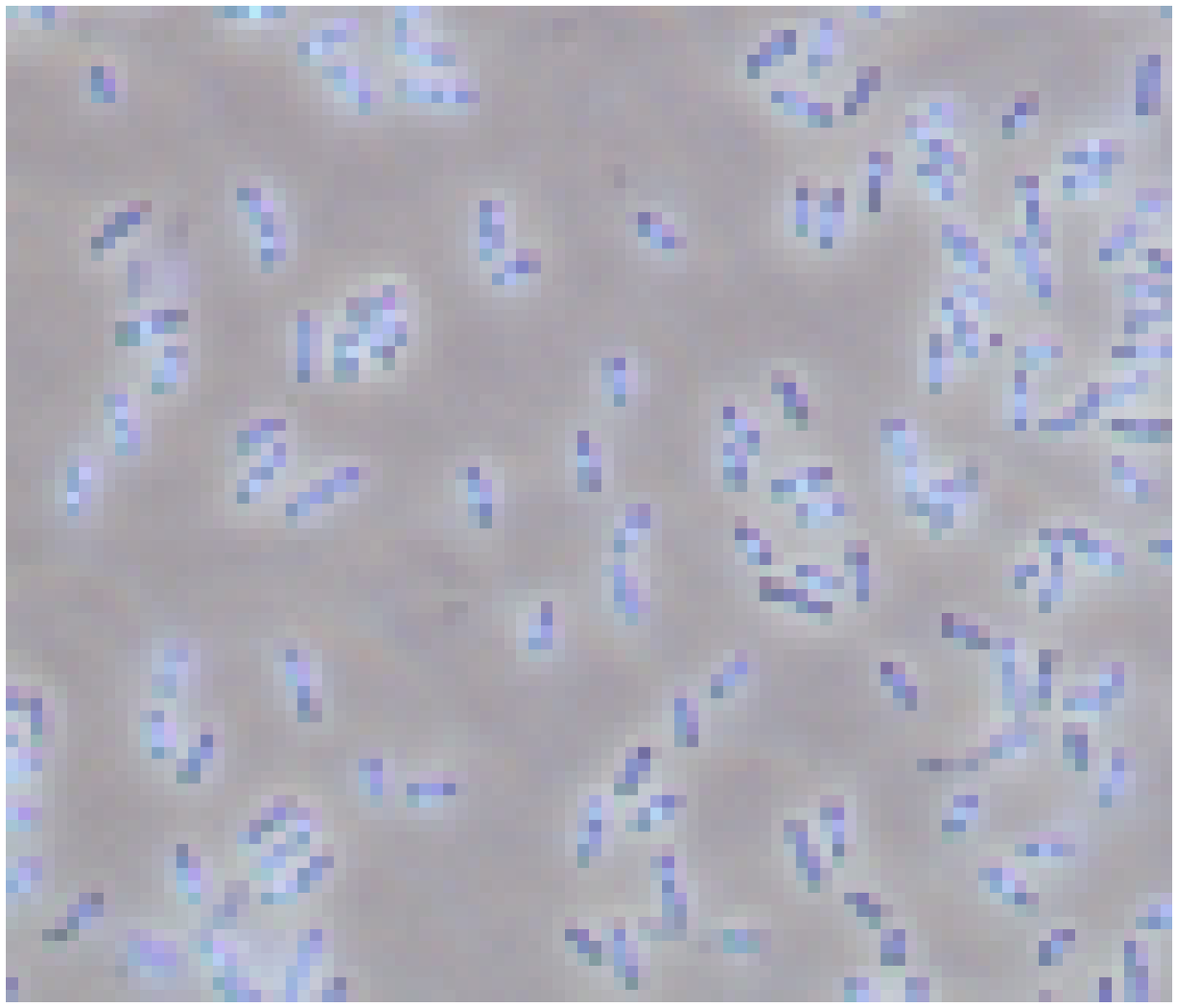}&
\includegraphics[width=.25 \textwidth]{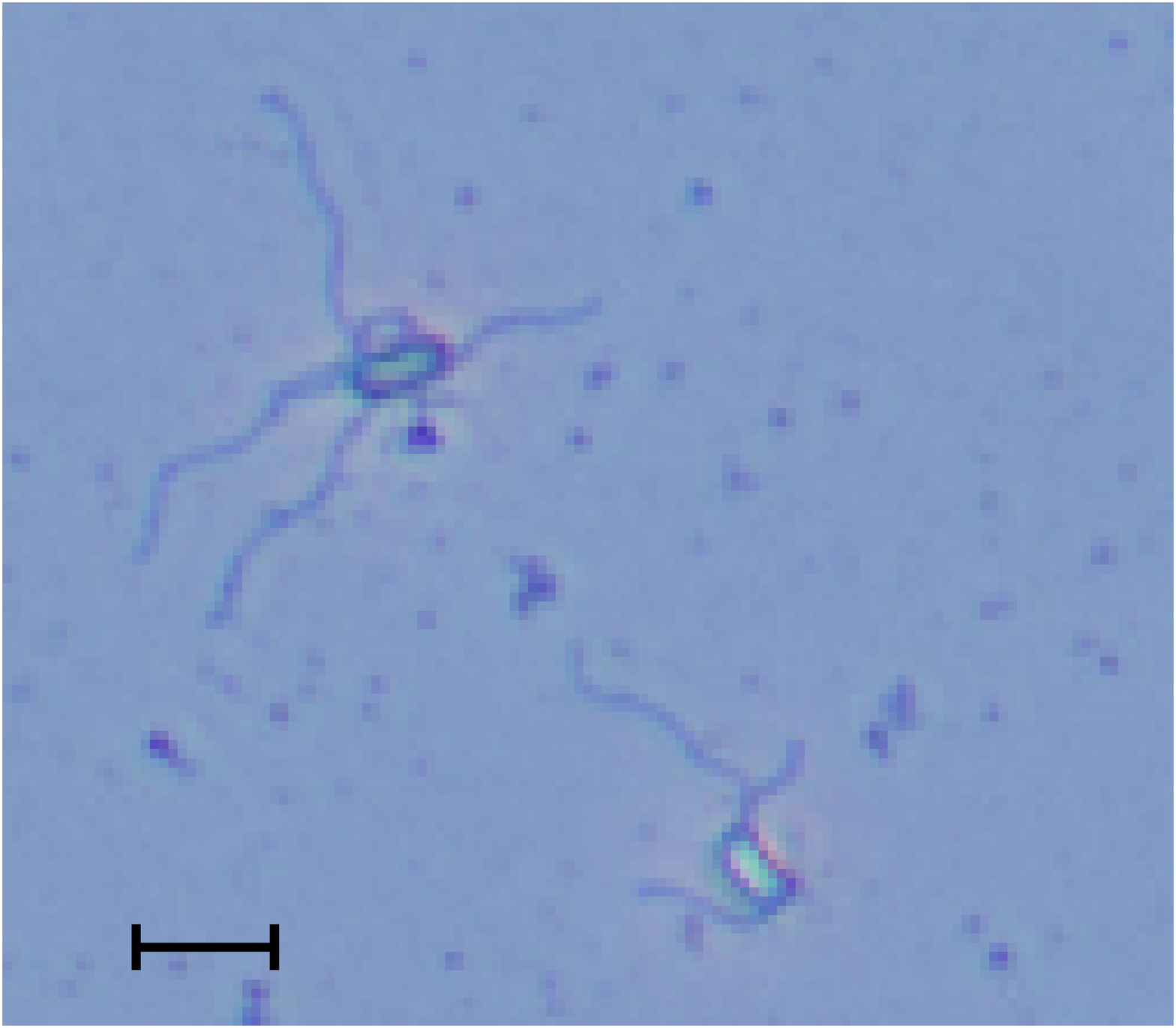}
\end{tabular}
 \caption{\label{cells} (a)  (Color online) Mononucleated {\sl E. coli} MG1655 swarmer cells,
   stained by DAPI fluorescence dye and visualized by phase contrast
   microscopy. (Bar=$5\ \mu $m). (b)   (Color online) Flagella of swarming
   cells visualized with the Saizawa and Sugawara procedure
   \cite{Harshey}. (Bar=$5\ \mu$m).}
\end{figure}

The development of an {\sl E. coli} MG1655 colony on a semi-solid
surface ($0.5\%$ "Difco" Agar) under certain nutritional (1$\%$ 
"Difco" Tryptone, 0.5$\%$ "Difco" yeast extract, 0.5$\%$ NaCl and $0.5\%$  D-(+)-glucose) 
and environmental (at $37^{\circ}$C and on average $22\%$
relative humidity) conditions has been studied in detail with 
round and square laboratory plates of $9$ and $12$ cm
respectively. We inoculated a $2.5\ \mu
l$ liquid drop of stationary phase bacteria cultivated overnight (in
LB, \cite{GG}, at $37^{\circ}$C). The following experimental 
observations were made:
\begin{itemize}
\item{The colony expands on the surface in a compact and round shape with
    few (typically 6 to 8) wide branches. The colonization of the plate takes place very
  rapidly: at
  $37^{\circ} $C a plate of 9 cm diameter is
  colonized in roughly one day,  see Fig.\ \ref{Ecoliflor}(a).}

\item{It was observed that the fronts of two colonies inoculated on
the same plate  never
merge unless they meet tangentially.  Branches do not simply grow radially from
their inoculation point, but avoid this intercolonial demarcation region, see
Fig.\ \ref{Ecoliflor}(b).}

\item{The term ``swarming'' is used in the literature to describe the active, flagellar-dependent
  surface motility of bacteria \cite{Harshey}. To confirm the
  relevance of flagella in the  mechanisms of surface
  translocation of this colony we have
  constructed by P1 transduction \cite{miller} a flagellar-defective
  mutant, MG1655 {\sl flhD}::Kan \cite{pruss}. The {\sl flhDC} operon encodes a key master regulator
  of the hierarchical system controlling  the synthesis of the
  bacterial flagellum in {\sl E. coli} \cite{fla-}. Having this operon inactivated the strain  MG1655
  {\sl flhD}::Kan can thus not synthesize the bacterial flagella. The
  wild type and the flagellar-defective mutant were inoculated
  simultaneously. After one day the mutant had initiated expansion
  over the surface, due to the
  combined influence of the expansive forces of bacterial growth
  (population pressure) and the
  fluid production.  But, in contrast with the wild type MG1655, it failed to
  continue the rapid surface colonization, see
  Fig. \ref{Ecoliflor}(c). The slow passive form of surface translocation
  displayed by the mutant is called ``spreading'' \cite{Matsuyama}. The same pattern was
  also observed when the flagellar-defective mutant was
  inoculated alone on a plate and for lower agar concentration (data
  not shown).  Notice also the existence of a clear asymmetric intercolonial demarcation region which
has been avoided by the fast expanding wild type colony MG1655.  This
  experiment shows thus that
  flagellar motility is essential for the surface motility of MG1655
  and that we are dealing with an active process of flagellar-dependent swarming
  motility that is able to {\em avoid} a region. 

}

\item{The macroscopic characteristics of the experimental
    patterns are extremely sensitive to humidity which in turn limits
    the fluid production. When the fluid production is high the pattern is
  almost compact with circular envelope and has few wide branches with cells deposits and 
    channels structures within. Whereas when the fluid
  production is limited the colony expands irregularly, with long,
  relatively thick branches with high
    cell densities (see Fig.\ \ref{Ecoliflor}(d)), very few secondary branches and greater
  gaps (see Fig.\ \ref{results}(b)(d) and (f)). 
}

\item{There is a transparent surrounding envelope of slime produced by the colony,
    see the lower arrow in Fig.\ \ref{Otros_Ecoliflor}(a), and channel-like
   structures
    within the colony surface, see the upper arrow in
    Fig.\ \ref{Otros_Ecoliflor}(a).}

\item{Along the edge in contact with the slime, one can observe a thin layer of
     quiescent cells apparently packed in an ordered way. See the lighter line of packed swarmer cells in
    Fig.\ \ref{Otros_Ecoliflor}(b), pushed by the turbulent collective motion
    of groups of swarmer cells in the interior region.}

\item{  In this swarming pattern we found, both in the center
and the periphery of the pattern, mononucleated
active cells, see Fig.\ \ref{cells}(a), from $3$ to $6\ \mu $m and
also with 4 to 6 flagella, see
    Fig.\ \ref{cells}(b). The swimmer cells of {\sl E. coli} MG1655 found in other patterns with
interstitial submerging are roughly $2\ \mu $m long with 4 to 6
flagella (data not shown). Therefore, under these experimental conditions the swarmer
cells of {\sl E. coli} MG1655 do not seem to undergo a significant change in their
cellular characteristics for surface translocation.}


\end{itemize} 

In order to choose an appropriate
   model certain features must be emphasized from the experimental
   observations:   In spite of the small, non-hyperflagellated
     swarming cells, the colony expands very rapidly on the agar showing a
   strong sensitivity to the surface humidity conditions, we therefore conclude
     that the lubricating fluid is relevant for the successful
     expansion of the colony.  Furthermore, we
     believe that the channel-like three dimensional structures are accumulations
     of bacteria in regions with higher concentration of fluid. As for the intercolonial demarcation region, our hypothesis is that the motion of motile cells is driven by
negative chemotaxis avoiding the regions with  high
concentrations of cells and the substances produced by them. Notice in
   particular the wide intercolonial demarcation zone in
   Fig. \ref{Ecoliflor}(c). Thus despite the fact that the mutant
   colony continues to grow (notice the long bacterial filament) which
   indicates the presence of
   nutrient in this zone, the swarming wild type colony {\em avoids} this
   region. This suggests that the demarcation zone cannot solely be due to
   nutrient depletion and there must be some chemotaxis in response
   to chemical signaling, nutrient level, pH variations, etc.

There are a few other reported examples of colonial
pattern formation in the literature, most of them imply the splitting
formation of tips and
branches but lead to different patterns since the mechanisms of
advance and self-organization are also different. We propose
 to describe this dynamical system by means of a reaction-diffusion
system, a standard tool in the field of bacterial colony pattern
formation, including well known and widely used features such as
lubrication and chemotaxis. The collective diffusion term
of the colony must account for the new self-organization mechanisms
observed in our experiments: we have seen that an increased density of cells and
slime in the tips of the colony helps the colony to advance, being less
sensitive
to local irregularities and showing thus, in conditions of adverse
motility, a poorly branched pattern.

\section{Reaction-diffusion Modeling}\label{model}

 Previous studies on pattern formation of other lubricating bacteria such as {\sl
  P. dendritiformis} var. {\sl dendron} have
    confirmed that reaction-diffusion models with a lubricating fluid
and chemotactic motion  can successfully describe the development of branching structures
    \cite{Bjacob}. However, when the colonial motility is decreased these models predict a  pattern transition 
    from a compact to a highly
   ramified pattern with thin branches, which is different from the one observed
  here. 

We will next construct a mathematical model of the reaction-diffusion
type, to describe our experimental observations.
The cells in the colony can be in a normal
static phase $S$ or in a differentiated motile state $V$ where the cells
produce liquid and use their flagella to
move.  Let $U$ be the nutrient consumed by cells in their reproduction, see Eq. (\ref{Uc}).  We include the fluid
substance $C$ that is excreted at a rate $\beta$ by motile  cells
and  disappears with time at a rate $\mu_c$ (being reabsorbed into the agar
medium or consumed or degraded by the cells), see Eq. (\ref{Cc}). The initial
condition is the circular
inoculum (with radius $R$) of a drop of liquid $C$ with bacteria $S$ in the static state in a
plate with nutrient $U_0$.  When
the concentration of fluid is sufficiently high, greater than a threshold $C_c$, these
static cells
reproduce at a rate
$\lambda$ into more static cells (in our model their reproduction rate is increased in
liquid media), see Eq. (\ref{Sc}). When the fluid level is too low, i.e. drops
below the critical value $C_c$, and bacteria are in
contact with the surface the swarming behavior is triggered: the new
cells are motile cells $V$ that are able to produce
fluid and move, see Eq. (\ref{Vc}). With time, motile cells return to the
static state at a rate $\mu$.  The diffusion term in Eq. (\ref{Vc})
 can be expressed as $\nabla F$ with $F=V v_V$ a flux of
  motile cells of density $V$ and with velocity $v_V$ 
proportional to $V$, $C$ and the gradient of $V$. This velocity
represents the fact that motile cells move faster for higher concentrations of
motile cells $V$ and lubrication fluid
 $C$, and that the direction of their motion is actively
  defined by negative chemotaxis moving towards regions of lower
  motile cell densities $V$ and substances produced by them (this
negative  gradient term could alternatively describe a population
pressure in the case of passive translocation).
Let $D_v$ be
the diffusion parameter of motile cells. The
 nutrient and fluid fields also diffuse with diffusion constants $D_u$ and $D_c$ respectively. To introduce the local
irregularities of the agar we include a local random viscosity term
$\eta(x,y)=\xi$ (with $\xi$  a
random number of uniform distribution in the interval $[0,1]$) which is constant in time and affects the nutrient diffusion. Our adimensional model reads:

\begin{equation}\label{Vc}
\frac{\partial V}{\partial t} =  + \lambda USC \Theta_{(C_c-C)}- \mu V + D_v
\nabla \cdot \left[ V^2 C  \nabla V  \right],
\end{equation} 
\begin{equation}\label{Sc}
\frac{\partial S}{\partial t}=+\lambda USC \Theta_{(C-C_c)} + \mu V,
\end{equation}
\begin{equation} \label{Uc}
\frac{\partial U}{\partial t} =-\lambda USC  +  D_u  \eta(x,y)  \Delta U,
\end{equation}
\begin{equation} \label{Cc}
\frac{\partial C}{\partial t}=  + \beta V - \mu_c C +    D_c \Delta  C,
\end{equation}
where $\Theta$ is the Heaviside step function (other
 thresholding functions lead to similar results). On swarming experiments a
 lag phase
 is always observed after 
inoculation when bacteria proliferate by cell division
at the central spot without any migration. Once the colony has reached some
threshold, the swarming process starts, i.e. rapid
 surface migration is preceded by a cell density-dependent lag period
 \cite{Kearns}. The
 thresholding condition of our model allows us to describe the
 biological colonial switch between the swimming and swarming state
 and this aforementioned initial time lag
 between inoculation and colony spreading. 

We have solved this system numerically using a finite difference
  scheme and
  an alternating direction technique for the diffusion term. We integrate inside a
circular region of radius $112.5$
 and set the
initial conditions to
$U(x,y;t=0)=U_0$,  $V(x,y;t=0)=0$, 
$C(x,y;t=0)=\xi(x,y)\Theta_{[R^2-(x^2+y^2)]} $ and
$S(x,y;t=0)=\xi(x,y)\Theta_{[R^2-(x^2+y^2)]}$
 with
 $R=5$ and $\xi$  a
random number of uniform distribution in the interval $[0,1]$.  We fix 
$\mu=0.15,\mu_c=0.5,D_c=0.015,D_u=D_v=1.5$ and $C_c=1.5$ and let other parameters vary
depending on the experiment. 

\begin{figure}
\begin{tabular}{c}
 (a)  \\
\includegraphics[width=.35 \textwidth]{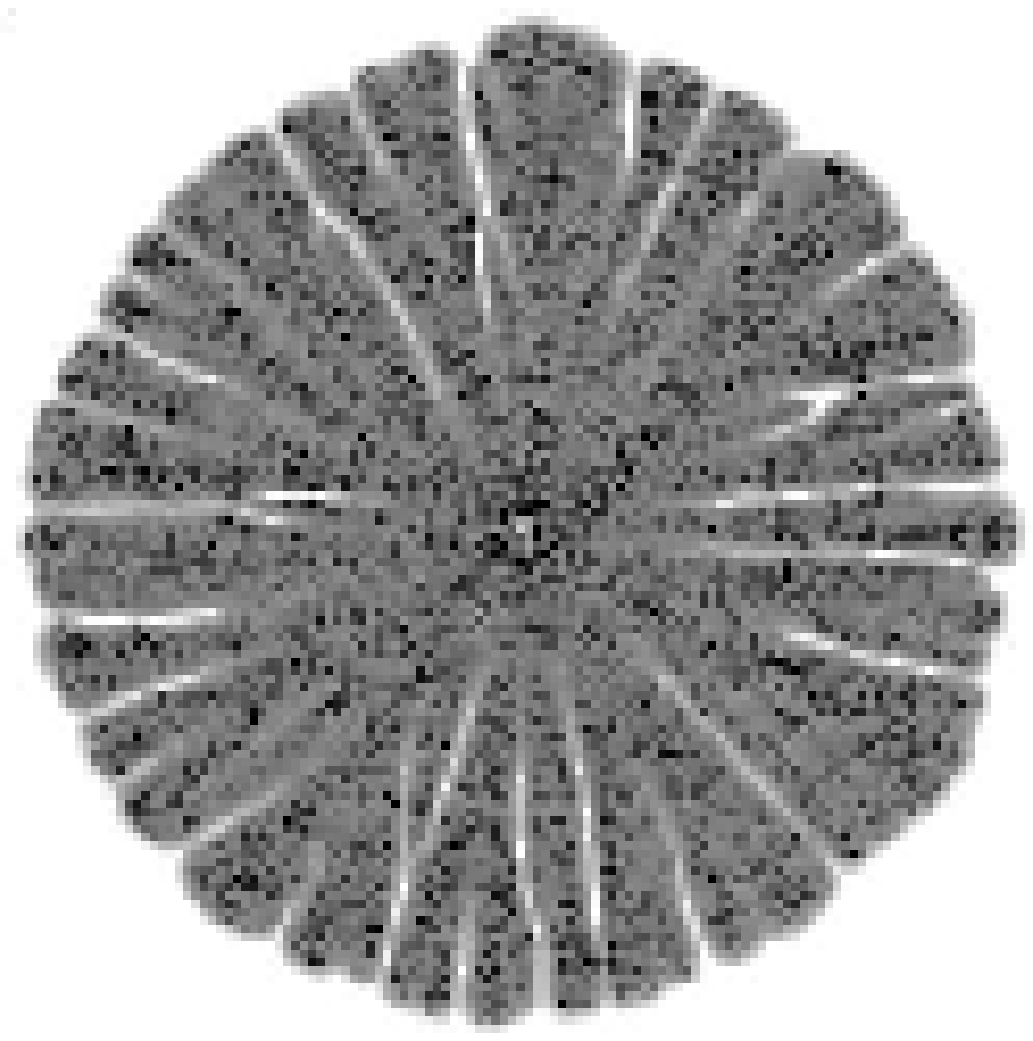}
\end{tabular}
\begin{tabular}{cc}
 (b) & (c)  \\
\includegraphics[width=.2 \textwidth]{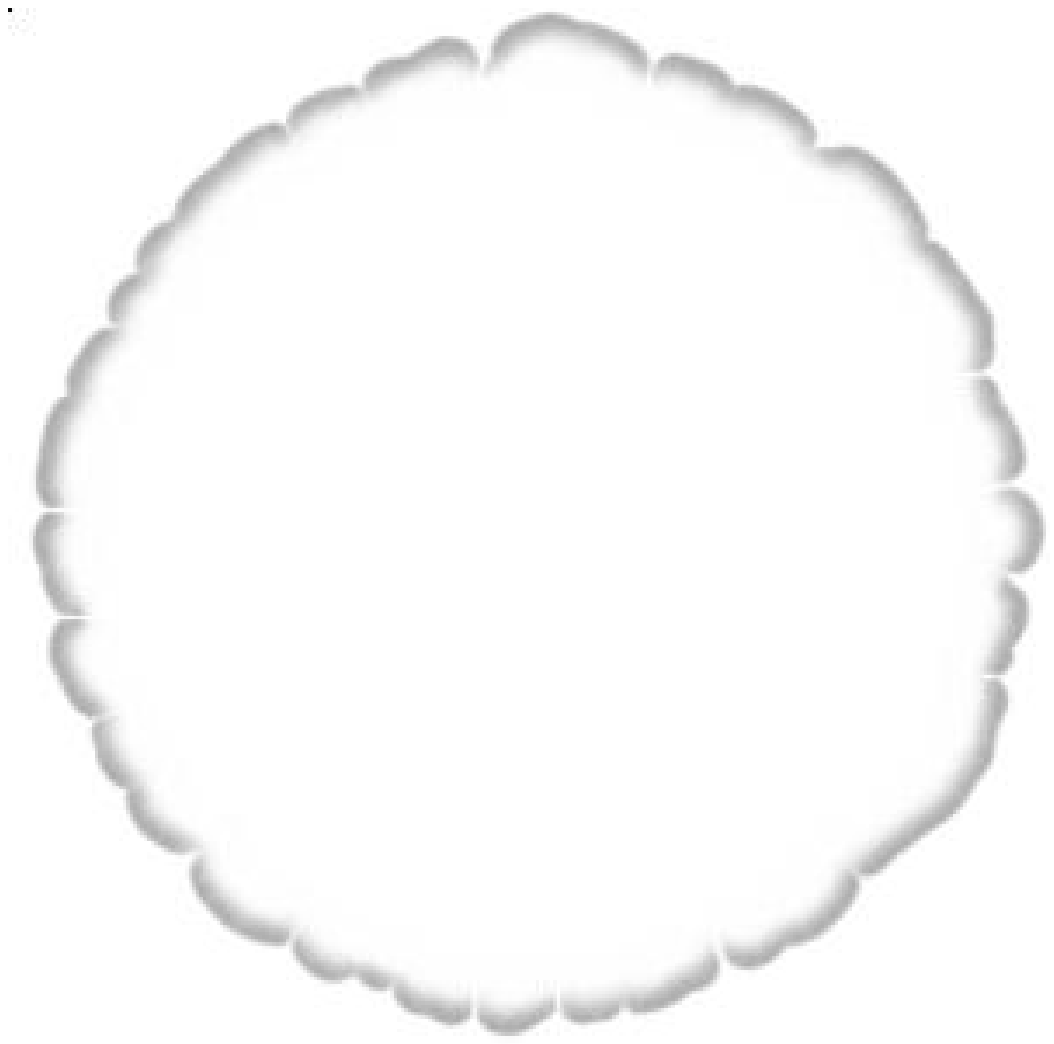}&
\includegraphics[width=.2 \textwidth]{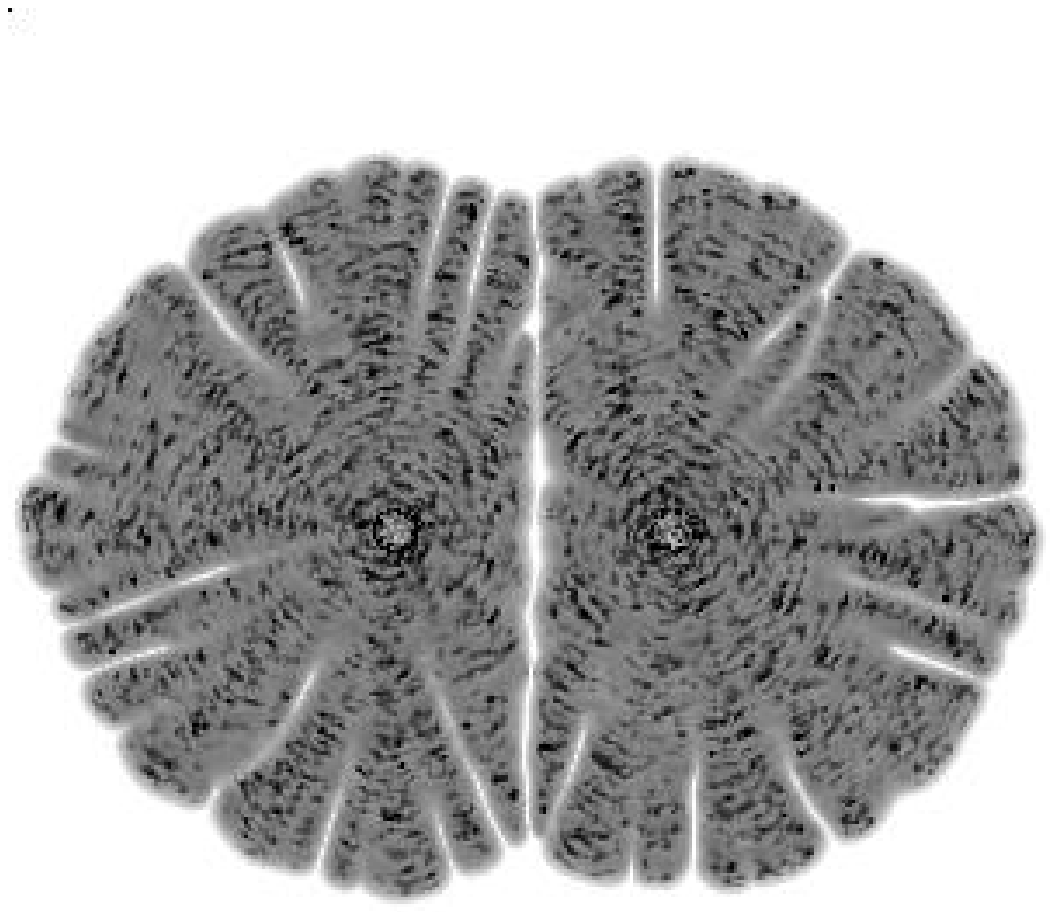}
\end{tabular}
 \caption{\label{NumEcoliflor} Numerical simulations results of reaction-diffusion growth with
   slime, when $\lambda=1$, $U_0=2$  and $\beta=1.3$. Local concentration of motile
   and static cells $V+S$ (a) pattern
   from single inoculation, shown at $t=650$, (b)  same study case,
   local density of fluid $C$, (c) pattern from double
   inoculation, shown  at $t=465$.}
\end{figure}

If we set  $\lambda=1$, $U_0=2$  and $\beta=1.3$ we
reproduce the experimental pattern. Figures \ref{NumEcoliflor} (a) and (c)  display the local bacterial
density $V+S$ at a certain time for one and two inoculi respectively. The
color coding ranges between white for
zero density and black for values equal to or higher than
$5$. The inner black
structures correspond to regions where temporary high concentrations of fluid
led to higher reproduction rates and an increment in the static
bacterial density. The
demarcation region induced by the gradient of the diffusion term can also be distinguished. For the study case (a), we show the local density
of fluid $C$ at the same given time, see Fig.\ \ref{NumEcoliflor} (b). This
fluid will disappear with time (it is reabsorbed into the agar or used by the bacteria
behind the front). 

\section{Numerical results and experimental validation}\label{val}

\begin{figure}
\begin{tabular}{ccc}
  (a) & (b) \\
\includegraphics[width=0.2 \textwidth]{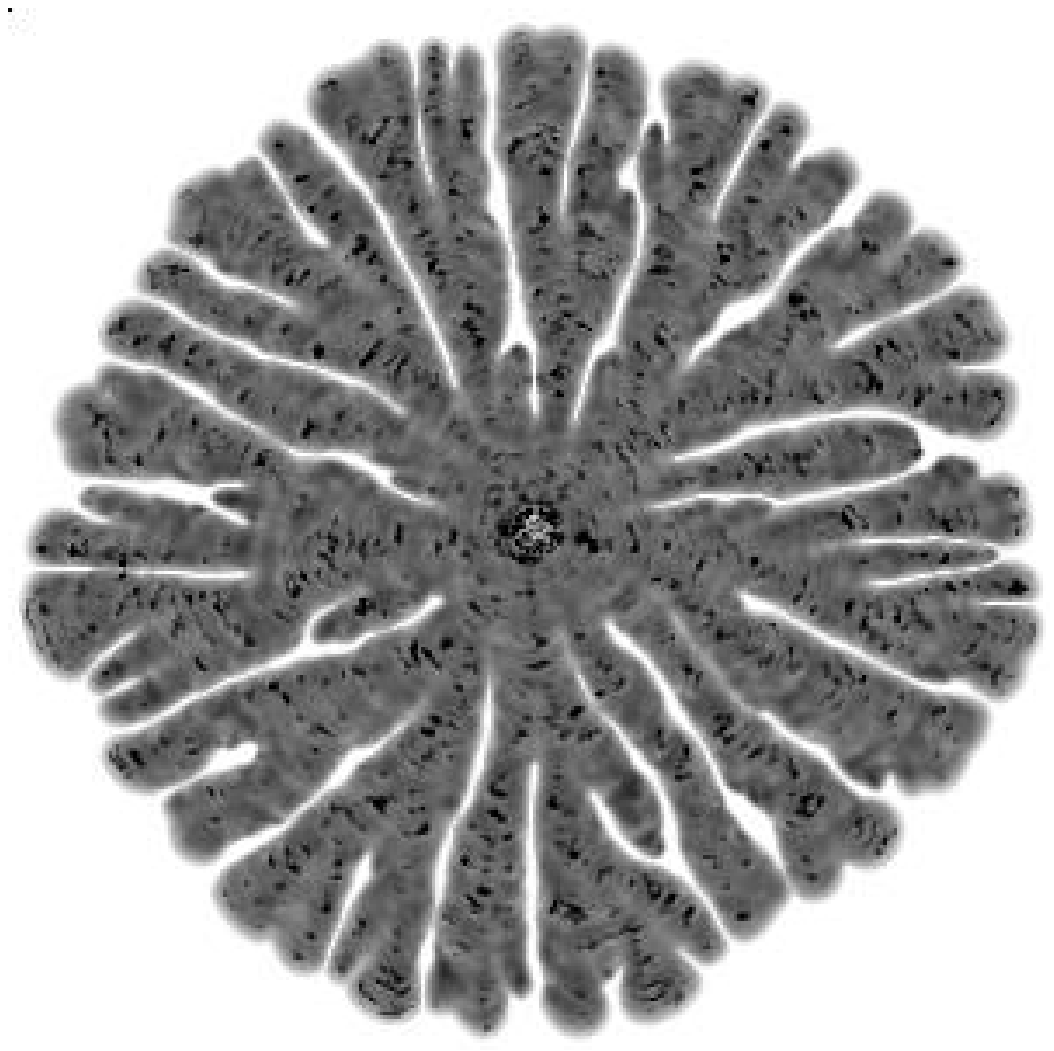}&
\includegraphics[width=0.2 \textwidth]{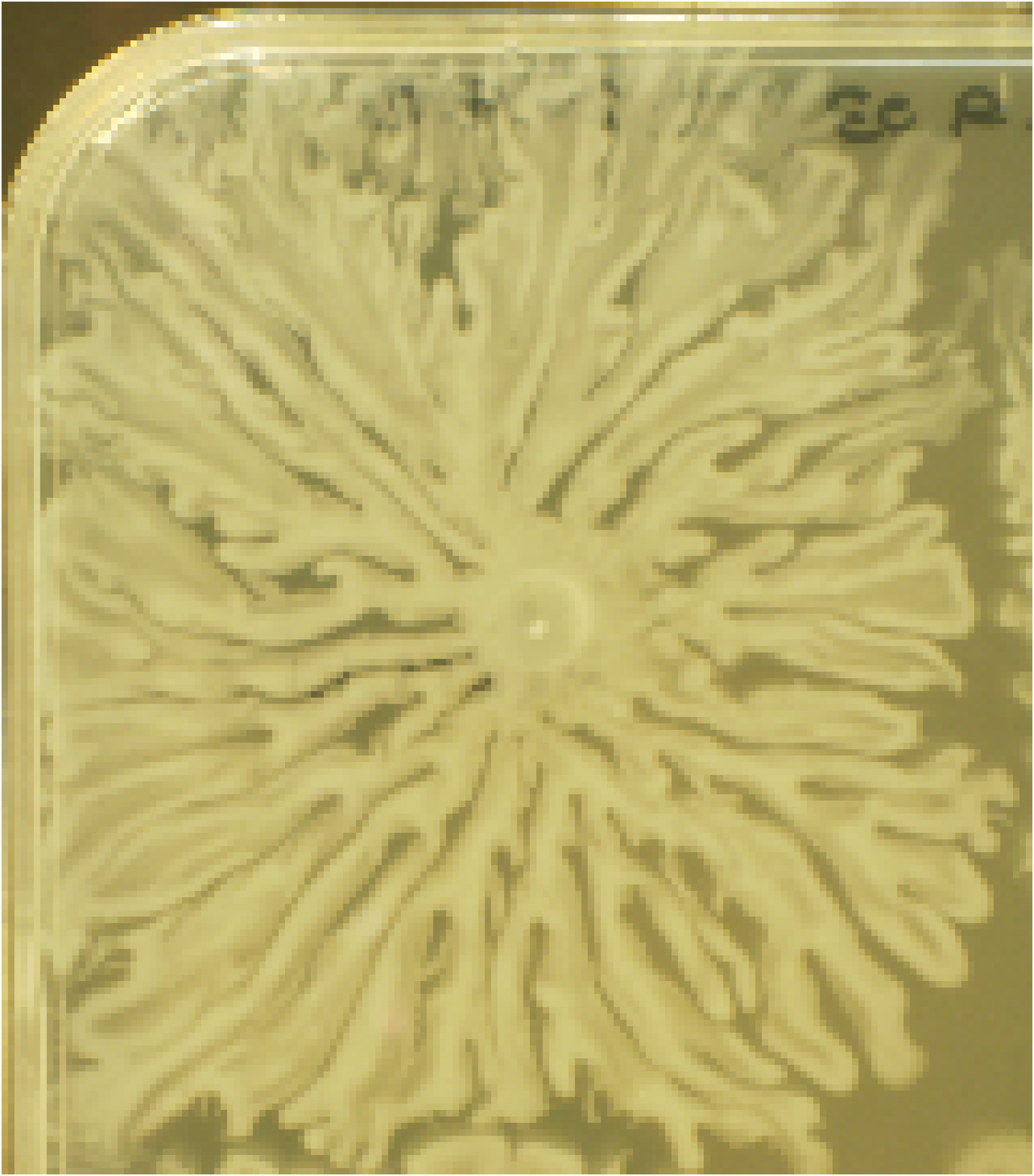}\\
  (c) & (d) \\
\includegraphics[width=0.2 \textwidth]{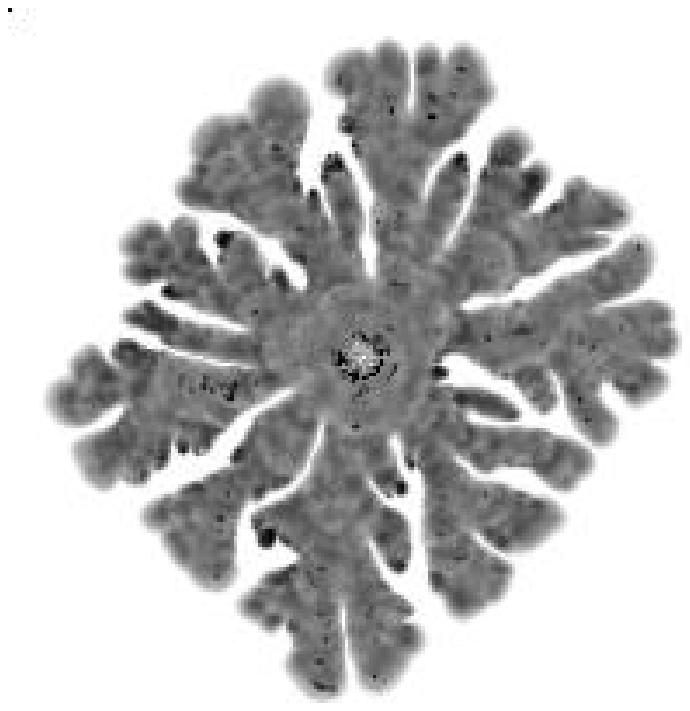}&
\includegraphics[width=0.2 \textwidth]{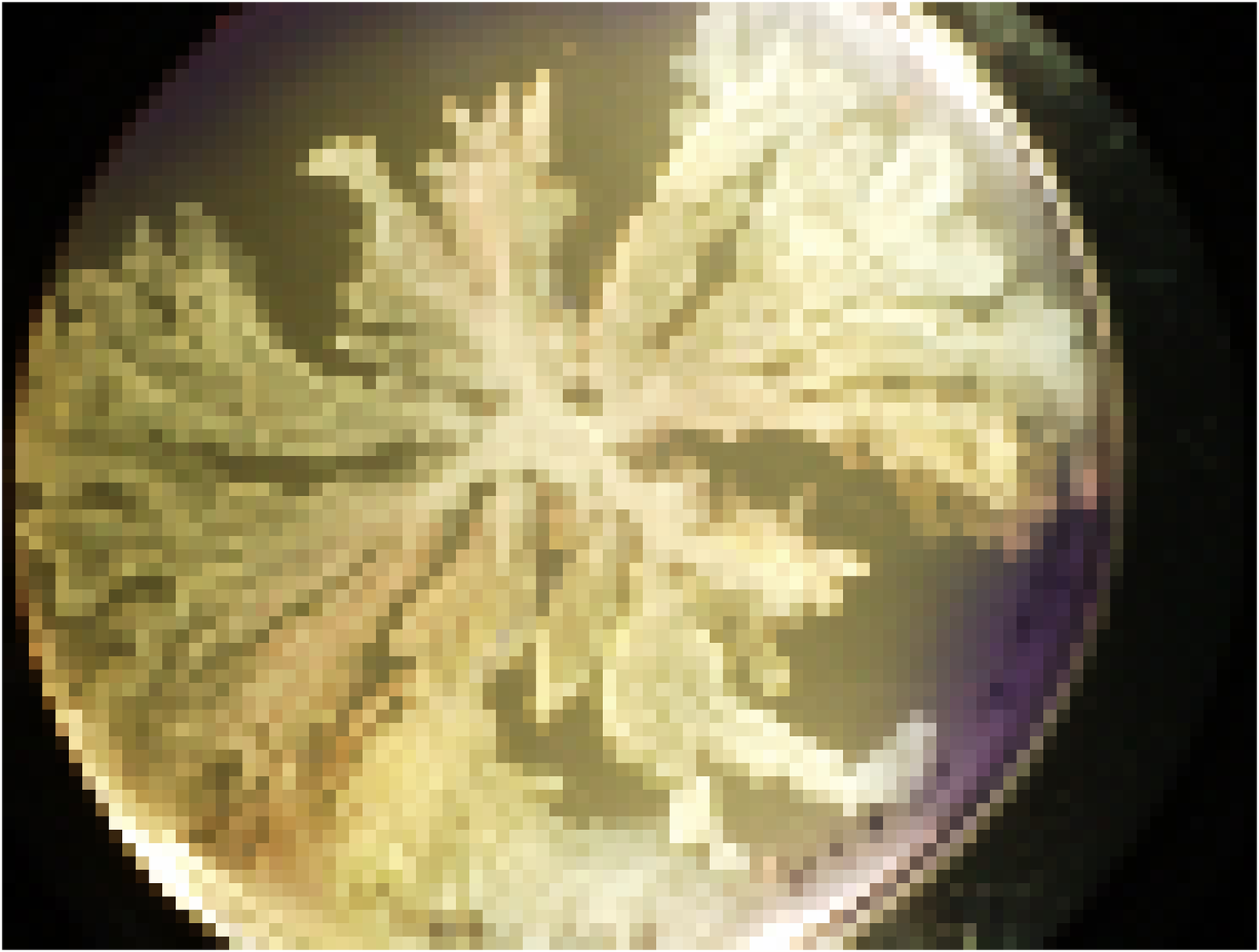}\\
  (e) & (f) \\
\includegraphics[width=0.2 \textwidth]{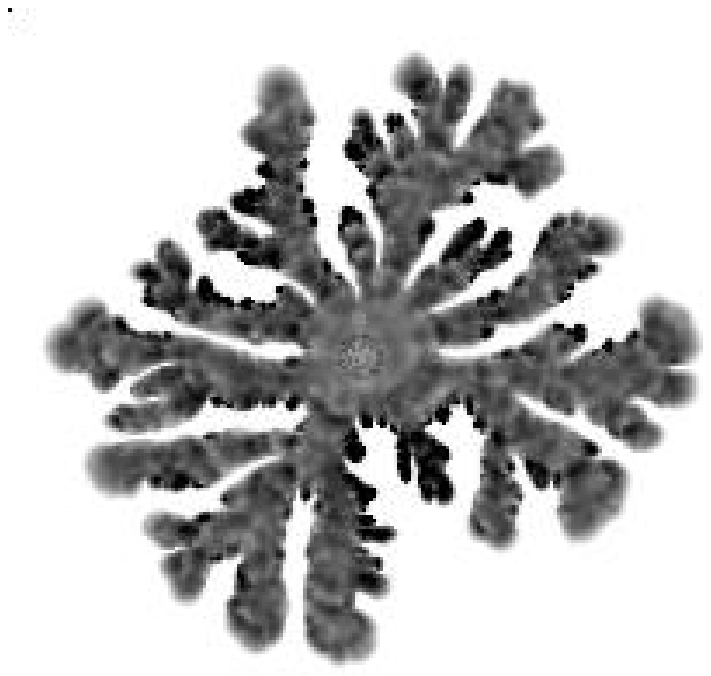}&
\includegraphics[width=0.2 \textwidth]{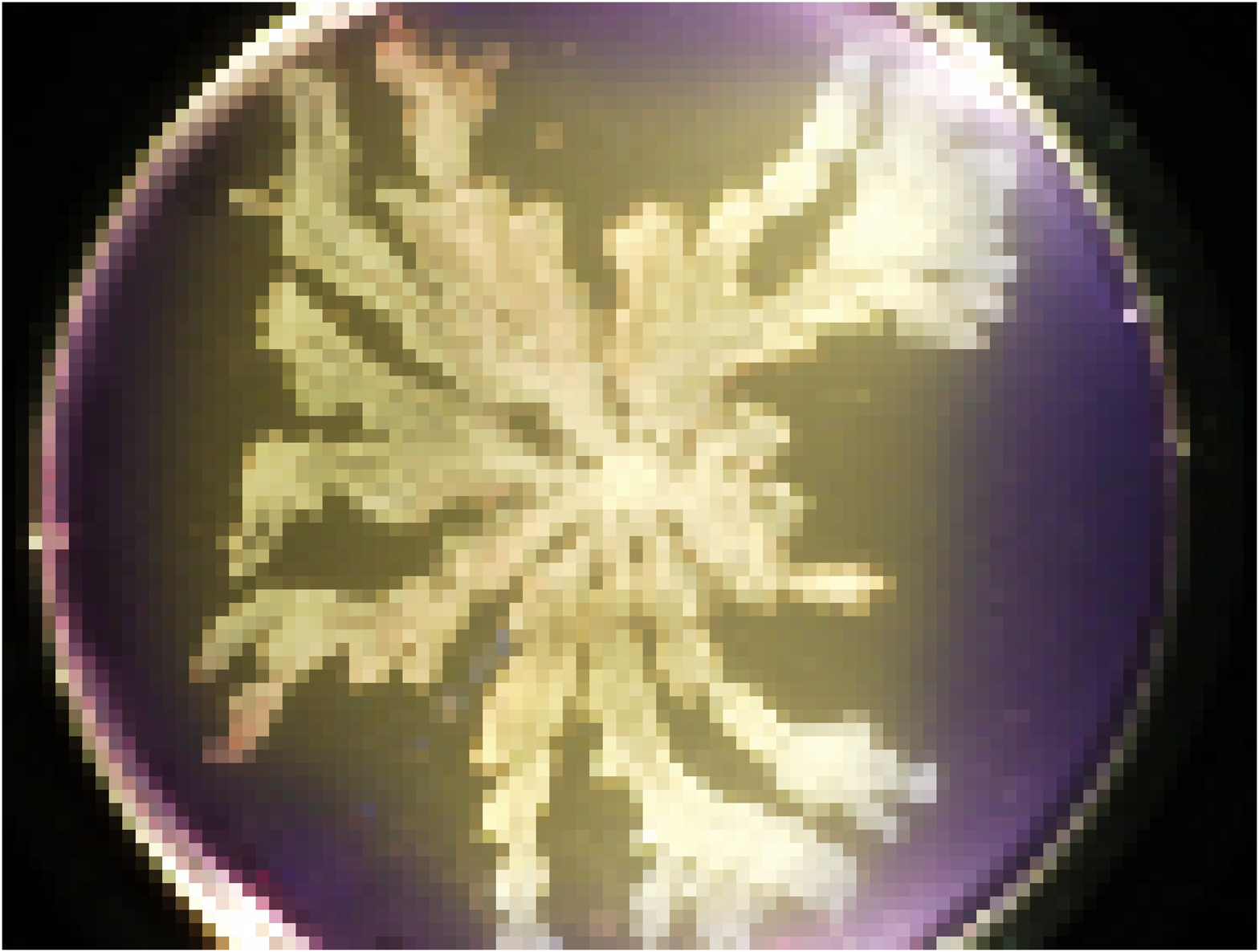}
\end{tabular}
\caption{\label{results}Numerical (left) and experimental (right) results when the production of fluid is limited. (a) $\lambda=0.6$,
  $U_0=2$, $\beta=1.3$ at $t=875$  (b)  (Color online) colony of size $6\times6$ cm$^2$
  approximately,  grown on a square plate with $0.5\%$ glucose at lower temperature ($30^{\circ} $C) (c) $\lambda=1$,
  $U_0=1.5$, $\beta=1.3$ at $t=600$  (d) (Color online) 
  grown with $0.3\%$ glucose at $37^{\circ} $C (e)  $\lambda=1$, $U_0=1.5$, $\beta=1$ at $t=750$
  (f)  (Color online) grown with  $0.3\%$ glucose at $37^{\circ} $C, when the surface lost $1\ g$ of
  water due to direct heating.}
\end{figure}

The model predicts a pattern transition  from compact to branched growth as the colony motility is limitted due to a
reduction in any of the factors in the collective diffusion term in
Eq. (\ref{Vc}): $D_v,V$ or $C$. For
instance if the reproduction rate ($\lambda$), the nutrient content ($U_0$) or the
possibility to produce slime using water from the agar ($\beta$) are decreased
separately the pattern becomes less compact and with an irregular
envelope, with relatively thick branches and very rare secondary branches. These numerical results are shown in
Fig.\ \ref{results}(a),(c),(e) and compared with experimental results where
the collective motility is reduced by lowering the temperature (b), the
glucose content (d) or the surface humidity (f) respectively.
Experimentally the most critical parameter was the humidity:
variations of only $5\%$ of the surface humidity had a significant
impact on the emerging
patterns.


Although the simple pattern analogy is not sufficient to
conclude that the model is correct, predicting the behavior when
several control parameters are changed is a good tool to validate
it. Under the circumstances reported here the parameter that had the strongest
impact on the emerging pattern was indeed the humidity. But  under different
experimental conditions other factors, such as local cell densities and quorum
sensing or type and quantity of nutrient supply required for the
flagellar synthesis and activation, may be
considered as essential.

\section{Conclusions}\label{conclu}

Branching is due to the sensitivity of the system to local
irregularities whose small perturbing effect is enlarged due to the
non-linearity of the diffusion term describing the cooperative
spreading of the colony. Branching patterns have also been seen in other bacteria such as
    {\sl B. subtilis}, \cite{Matshushita}, and in  {\sl
    P. dendritiformis} var. {\sl dendron} (genera
    Paenibacillus of {\sl B. subtilis}), \cite{Bjacob}. In these colonies, when the motility of the colony is strongly limited, the advancing front is
extremely sensitive to local irregularities and shows very frequent tip
splitting leading to very ramified, thin branches.  In some cases
(like the so-called Dense Branching Morphology) the distribution of
  branch lengths is even exponential, suggesting that the tip-splitting of branches takes
place at random \cite{Matshushita}.

In our study case
when the motility is reduced, the density of motile lubricating cells
in the advancing front is incremented, the fluid level increases also
and the advancing branches are thick, with low branching rate. 
This mechanism of response of the colony to adverse conditions has been
  implemented mathematically as follows: a thresholding condition on
  the fluid level launches the production of lubricating swarming
  cells V and the reproduction rate is increased in fluid rich regions. Since the collective colonial 
motility expressed by the diffusion term shows a strong dependence on
the density of motile cells $V$, the
very dense branches are 
less sensitive to local irregularities and show thus a lower branching
rate.

Studies such as this one are aimed at understanding the principles underlying
self-organized pattern formation processes and establishing
a systematic correspondence between the observed macroscopic colonial
patterns and the putative microscopic collective physics. The new
biological pattern has indeed given us some hints about the underlying
collective physical principles or colonial strategy. 

This strategy, or
part of it, may be common to other bacterial patterns. For instance branching patterns have been reported recently
  on a colony of {\sl B. subtilis} where potassium ions were
  shown to increase the production of lubricating fluid with surfactin
  and thus
 increment the motility substantially without formation of
  flagella (not swarming)\cite{Kinsinger}. When the
  motility is reduced due to a low fluid production the colony shows
  also thick branches with relatively low
 branching rate. A fluid preceding each dendritic finger of growth
 can be seen. From analysis of this information alone, we conjecture that a similar mechanism may take place in this case: under adverse conditions the production of lubricating cells would be increased, reproduction may also be more effective in fluid rich
 regions and in turn these branches would have higher density showing again a low
 sensitivity to local irregularities and low branching rate. The
 population pressure would also induce here a negative gradient term.

Swarming in {\sl E. coli} has been previously reported on other
strains that swarm due to a differentiation from the swimming
state to a swarming-adapted strongly
hyperflagellated, multinucleated state \cite{HarsehyColi}. This
present work reports on the swarming ability of
another {\sl E. coli} strain, the wild type MG1655, that shows a greater variety of patterns. Instead of cell
hyperflagellation, the strategy chosen for colonization is a strong
fluid production and incremented density of motile cells, the result is a
fast advancing colony with low branching rate. Since  {\sl E. coli} is a
genetically well characterized organism, the
experiments reported here can be used for further studies of the genetic expression associated to the collective,
self-organized, swarming pattern formation process.





\section{Acknowledgments}
We thank L. V\'{a}zquez,  A. Giaquinta and J. Shapiro for support and
careful reading of the manuscript. We are grateful to Birgit
M. Pr\"{u}\ss\ for providing an {\sl E. coli} strain with the
{\sl flhD}::Kan allele and to Mary Berlyn (E.coli
Genetic Stock Center) for suppling the {\sl E. coli} MG1655
(\#CGSC6300) strain. 
Special thanks go to J.-A. Rodr\'{\i}guez-Manfredi and F. Camps-Mart\'{\i}nez for their
invaluable assistance with images and videos. The work of M.-P. Zorzano, J.-M. G\'{o}mez-G\'{o}mez and
M.-T. Cuevas is supported by a INTA fellowship for training in Astrobiology.

\end{document}